\documentclass[12pt]{article}

\usepackage{graphicx}
\usepackage{citesort}
\usepackage{mathbbol,amsmath,amssymb,amsfonts}
\usepackage{feynmf}

\newcommand{\beq}{\begin{equation}}
\newcommand{\eeq}{\end{equation}}

\newcommand{\beqa}{\begin{eqnarray}}
\newcommand{\eeqa}{\end{eqnarray}}

\newcommand{\bean}{\begin{eqnarray*}}
\newcommand{\eean}{\end{eqnarray*}}
\newcommand{\ra}{\rightarrow}
\newcommand{\da}{\dagger}
\newcommand{\pa}{\partial}
\newcommand{\ka}{\kappa}
\newcommand{\mc}[1]{{\mathcal{{#1}}}}
\newcommand{\ti}[1]{{\widetilde{{#1}}}}

\newcommand{\fs}[1]{\hbox{{$#1$}\llap{$/$}}}

\newcommand{\pv}{{\bf p}}

\begin{document}

\begin{center} 
{\Large \bf A Field Theory Model With a New \\ Lorentz-Invariant 
Energy Scale}
\end{center} 
\vskip .2cm   
  
\renewcommand{\thefootnote}{\fnsymbol{footnote}}   
\centerline{\bf Tomasz 
Konopka\footnote{tkonopka@perimeterinstitute.ca}}   \vskip .4cm  
\centerline{ \it University of Waterloo, Waterloo, ON, Canada, 
and } \centerline{ \it Perimeter Institute for Theoretical 
Physics, Waterloo, ON, Canada}

\begin{abstract}

A framework is proposed that allows to write down field theories 
with a new energy scale while explicitly preserving Lorentz 
invariance and without spoiling the features of standard quantum 
field theory which allow quick calculations of scattering 
amplitudes. If the invariant energy is set to the Planck scale, 
these deformed field theories could serve to model quantum 
gravity phenomenology. The proposal is based on the idea, 
appearing for example in Deformed Special Relativity, that 
momentum space could be curved rather than flat. This idea is 
implemented by introducing a fifth dimension and imposing an 
extra constraint on physical field configurations in addition to 
the mass shell constraint. It is shown that a deformed 
interacting scalar field theory is unitary. Also, a deformed 
version of QED is argued to give scattering amplitudes that 
reproduce the usual ones in the leading order. Possibilities for 
experimental signatures are discussed, but more work on the 
framework's consistency and interpretation is necessary to make 
concrete predictions.

\end{abstract}

\renewcommand{\thefootnote}{\arabic{footnote}}
\setcounter{footnote}{0}

\section{Introduction \label{s_intro}}

There is a large discrepancy between the physical pictures 
painted by standard quantum field theory on the one hand and, on 
the other, candidates for fundamental theories of nature such as 
loop quantum gravity or string theory. One describes the world in 
terms of particles on a smooth background, while the others 
involve much stranger objects like spin-networks, strings, or 
membranes. Moreover, the realms of validity of standard field 
theories and quantum gravity theories are separated by many 
orders of magnitude. The situation prompts many questions as to 
what kind of physics governs the intermediate energy scales. 

Dimensional arguments suggest that quantum gravity effects should 
become important at energy scales close to the Planck energy, 
$E_{Planck} = \sqrt{hc/G} \sim 10^{22}$ MeV. For particle 
accelerator experiments, this is an extremely high energy. 
Nevertheless, it has been realized in recent years that 
experiments detecting ultra-high energy particles from 
cosmological sources may already be or may soon become sensitive 
to phenomena around the Planck scale \cite{testable}. It would be 
therefore both interesting and useful to have some models that 
could predict or reproduce the results of these experiments. The 
purpose of this paper is to formulate and study a model that 
could bridge the divide between standard physics and quantum 
gravity, and thereby serve to understand Planck scale 
phenomenology. The aim is to remain conceptually close to 
standard quantum field theory so that computations can be 
assigned meaningful interpretations, but to nevertheless consider 
some general features that are expected to arise from quantum 
gravity. 

A simple way to incorporate the Planck scale into a field theory 
is to add new terms to the standard model Lagrangian. Models 
constructed in this manner have been extensively studied and 
their free parameters have already been tightly constrained 
\cite{Kostolecky,Alfaro,LIVReview}. However, since such 
approaches usually break Lorentz invariance in their very 
formulation, they are at odds with the candidates for fundamental 
theories of quantum gravity such as loop quantum gravity or 
string theory.

A candidate for the resolution of the problem with breaking 
Lorentz symmetry is Deformed Special Relativity (DSR). In DSR, 
the momentum space transformations of special relativity are 
modified so that the energy (or momentum - the details are model 
dependent) of a particle is saturated at a particular level much 
like the velocity is bounded by $c$ \cite{DSRoriginals}. The new 
invariant energy scale of the DSR models is called $\ka$ in 
concordance to earlier work on quantum deformations of the 
Poincare algebra \cite{DSROld}. An important aspect of DSR is 
that the scale $\ka$ can be understood as setting a curvature 
scale in the momentum space. The curvature gives rise to one of 
the characteristic features of DSR that the action of boosts on 
momentum variables is non-linear. (Indeed, it is the 
non-linearity of the transformations that implements the 
invariance property of the scale $\ka$.) Interest in the DSR 
approach is enhanced by indications that it may actually descend, 
in a certain limit, from candidates for a full theory of quantum 
gravity \cite{loopy,DSRin2plus1}. 

In this paper, the core idea of DSR that momentum space could be 
curved rather than flat is used to deform standard quantum field 
theory. The goal of the work is to write a field theory that 
incorporates the scale $\ka$ in a Lorentz invariant way and that 
is at the same time relatively easy to calculate with. Although 
the motivation for this work comes from the DSR program, the 
resulting model is distinct from constructions already proposed 
\cite{Daszkiewicz,Arzano,Dimitrijevic}. In fact, the approach in 
this paper is designed to avoid many of the complications, such 
as non-linear and non-associative momentum addition rules or 
non-commutative geometry, that arise in the other works. The 
exact relation between the proposed model and the other ones 
based on orthodox DSR is not addressed in detail and remains to 
be understood.

In the next section, the idea of curved momentum space is 
explained in more detail. Very importantly, it is pointed out 
that some well-known features of special relativity can be used 
as a guide in formulating a theory on (anti-)de Sitter momentum 
space. The discussion suggests to use framework that relies on an 
auxiliary dimension but characterizes physical particle states by 
a new constraint. Effectively, the proposal is to treat the new 
energy scale in much the same way as the first energy scale 
characterizing quantum field theories, the mass, is dealt with. 
In section \ref{s_scalar}, the formalism is used to define a 
deformed $\phi^4$ scalar field theory. The momentum space 
constraints are implemented at the level of the field expansions, 
interactions are defined through Feynman diagrams, and scattering 
amplitudes are obtained via a new set of Feynman rules. The 
deformed theory is shown to be unitary to second order in the 
coupling. In section \ref{s_QED}, the deformation procedure is 
applied to quantum electrodynamics. Properties and consequences 
of having momentum conservation in the extra dimension are 
discussed. Finally, section \ref{s_discussion} reviews the 
positive as well as the negative features of the proposed 
deformed field theories.


\section{Curved Momentum Space \label{s_curved}}

The structure of momentum space plays an important role in 
classical and quantum mechanics. It is usually assumed to be a 
flat, four dimensional manifold. In trying to introduce a new 
invariant scale into quantum field theory, it is tempting to 
implement it as a curvature in the particle momentum space. This 
idea is in fact the corner-stone of Deformed Special Relativity 
(DSR) \cite{DSRreview}, although it has already been explored 
earlier in other contexts \cite{Kadyshevsky}. The purpose of this 
section is to review the concept of curvature in momentum space. 
It is discussed in the context of both Deformed Special 
Relativity and standard Special Relativity, and the insights 
gained are used to propose an approach to field theory with an 
invariant energy scale based on an auxiliary dimension. 

A brief explanation of notation: Latin indices are 
three-dimensional ($i,j,k=1,2,3$), Greek indices are 
four-dimensional ($\mu,\nu=0,1,2,3)$, the metric on 
four-dimensional flat momentum space is 
$\eta_{\mu\nu}=\mathrm{diag}(+,-,-,-)$ so that $p^2=p_0^2-p_i^2.$ 
Some of the discussion refers to a large auxiliary dimension 
labelled by coordinates $p_4$. Any dependence on this extra 
dimension, as well as its signature, is always made explicit.

\subsection*{Deformed Special Relativity}
Although several reviews of the DSR formalism at varying level of 
technical detail are available \cite{DSRreview}, it is useful to 
briefly summarize some of its features here. It turns out that 
the momentum variables of DSR parametrize a four-dimensional 
space of positive curvature - a deSitter space 
\cite{Kowalski_deSitter}. A DSR model can be constructed from 
this insight as follows. Four-dimensional deSitter space can be 
viewed as the hyperboloid in a five-dimensional space with 
coordinates $(p_0,\, p_i,\, p_4)$ satisfying the constraint \beq 
\label{constraint} p_0^2 - p_i^2-p_4^2 = -\ka^2; \eeq the 
deSitter radius $\ka$ being the new invariant scale. It follows 
from the form of the constraint that it is invariant under an 
$SO(1,3)$ symmetry whose boost generators act as follows \beq 
\label{Pboosts} [N_i, p_0] = ip_i, \qquad [N_i, p_j] = 
i\delta_{ij}p_0, \qquad [N_i, p_4] = 0. \eeq Since the surface 
(\ref{constraint}) is four-dimensional, it is possible to 
describe it using a set of only four intrinsic coordinates. The 
map from the five embedding coordinates $(p_{\mu},\,p_4)$ to 
these new four-dimensional variables is not unique and different 
choices of the map lead to the many versions (bases) of DSR. For 
example, planar coordinates $\hat{p}_\mu$ related to 
$(p_{\mu},\,p_4)$ by \beq \label{planarcoordinates} p_0 = \kappa 
\sinh 
\frac{\hat{p}_0}{\kappa}+\frac{1}{2\kappa}\hat{p}_i\hat{p}^i 
e^{\hat{p}_0/\kappa}, \qquad p_i = \hat{p}_i 
e^{\hat{p}_0/\kappa}, \qquad p_4 = \kappa \cosh 
\frac{\hat{p}_0}{\kappa} - \frac{1}{2\kappa}\hat{p}_i\hat{p}^i 
e^{\hat{p}_0/\kappa} \eeq define the bicrossproduct basis. 
Because the map (\ref{planarcoordinates}) is non-trivial, the 
boost generators (\ref{Pboosts}) act non-linearly on the 
four-dimensional coordinates, \beq [N_i, \hat{p}_0] = i\hat{p}_i 
,\qquad [N_i, \hat{p}_j] = i\delta_{ij}\left( 
\frac{\ka}{2}\left(1-e^{-2\hat{p}_0/\ka}\right) 
+\frac{1}{2\kappa}\hat{p}_i\hat{p}^i\right) - 
i\frac{\hat{p}_i\hat{p}_j}{\ka}.\eeq A consequence of these 
commutators is that $\hat{p}_i \hat{p}^i$ is bounded from above 
by $\ka^2$, thereby imposing a frame-invariant momentum cutoff. 
Beside the algebraic structure, there is also a non-primitive 
co-product map $\Delta: \mc{P} \ra \mc{P}\otimes \mc{P}$ from one 
copy of the momentum algebra $\mc{P}$ to a tensor product. As a 
consequence of the non-primitive co-product, the space-time that 
is dual to the momenta $\hat{p}_\mu$ is non-commutative 
\cite{Majid}. Non-commutative geometry and the Hopf-algebraic 
structure of the bicrossproduct basis have therefore been the 
starting blocks of several approaches to constructing effective 
field theories in the context of DSR 
\cite{Daszkiewicz,Arzano,Dimitrijevic}. 

Current understanding of the proposed field theories and indeed 
of the whole DSR framework is still incomplete. Some of the 
important questions to be answered are related to how DSR should 
be interpreted as a physical theory and what its observable 
consequences are \cite{Ahluwalia,Liberati,GirelliLivine}. One of 
the ambiguities is due to the DSR literature making reference to 
several inequivalent momentum variables: the discussion thus far 
has mentioned the $p_\mu$ of the higher dimensional 
representation and the $\hat{p}_\mu$ of the bicrossproduct basis, 
but other variables are also possible. The different choices of 
coordinates are often useful for specific purposes. For example, 
a system of coordinates \cite{JudesVisser} in which conservation 
laws are linear has been used to study particle kinematics 
\cite{Heyman}. As another example, the embedding coordinates of 
(\ref{constraint}) has been used to define multiple particle 
states that do not suffer from the so-called `soccer ball 
problem' \cite{GirelliLivine}. Some of these coordinate systems 
have been related to each other and to the higher dimensional one 
\cite{Girelli:2005dc}, but there still remain questions as to 
which system should be used to construct field theories.

\subsection*{Special Relativity}

Despite the ambiguities in the interpretation of DSR, the concept 
of a curved momentum space is arguably as old as Special 
Relativity \cite{FloEteraSR}. In standard Special Relativity 
(SR), the mass-shell condition $p_0^2 - p_i^2 = +m^2$ can also be 
seen, similarly to (\ref{constraint}), to implement a curvature 
in the physical configuration space of relativistic particles. 
Interestingly, the curved momentum space of special relativity 
can be described using intrinsic three-dimensional variables, but 
such as formulation leads to non-commutative geometry just like 
in DSR \cite{FloEteraSR}. The physical manifestation of the 
non-commutativity, in the case of SR, lies in the non-linear 
addition law for particle velocities. In some sense, therefore, 
it is useful to think of standard special relativity as a 
lower-dimensional version of DSR, or, conversely, to think of DSR 
as a higher dimensional generalization of standard special 
relativity. 

There are several lessons to be learnt from comparing standard 
special relativity to DSR. One lesson is that the curvature in 
momentum space has physical consequences. In SR, that physical 
consequence is an upper bound on particle velocity and a velocity 
addition rule that is compatible with that bound. Another lesson 
is that, for practical purposes, it is easier to work with the 
higher-dimensional rather than the intrinsic variables. For 
example, calculations of scattering thresholds are easily done 
using four-momentum addition rules whereas these same 
calculations are much trickier in terms of the variables such as 
velocities that must be added in a non-trivial way. On a more 
philosophical level, yet another lesson to be learnt from 
standard special relativity is that higher-dimensional variables 
have a physical interpretation. As a result of SR, the unified 
approach to space and time (energy and momenta) is now 
well-established.

\subsection*{Higher-Dimensional Approach}

This work is aimed at writing field theories that incorporate an 
energy scale $\ka$ into their basic equations. The motivation for 
this work is taken from the DSR program which suggests to view 
four-dimensional momentum space as a constraint surface in a 
five-dimensional flat space. The approach taken draws on the 
lessons learnt from standard SR. 

For the purposes of this paper, a deformed-relativistic particle 
is defined in terms of a vector in flat five-dimensional momentum 
space subject to $\ka$-shell and $m$-shell constraints. The 
constraints are taken to be \beq \label{twoconstraints} 
\begin{split} p_0^2-p_i^2 &= m^2, 
\\  p_0^2-p_i^2-\xi p_4^2 &= -\xi \ka^2, \end{split} \eeq where the parameter 
$\xi=\pm 1$ is introduced to bring a little more generality than 
usually considered in DSR theories. The choices $\xi=+1$ and 
$\xi=-1$ correspond to choosing the $\ka$-shell constraint 
implement a de Sitter or anti-de Sitter momentum space. Since 
both equations are invariant under an $SO(1,3)$ symmetry, the 
resultant particle theory can be said to be Lorentz-invariant.

The mass shell constraint suggests to use standard terminology 
like `energy' and `momentum' to refer to $p_0$ and $p_i$, 
respectively. This terminology is different and should not be 
confused with that used in the DSR literature where these words 
refer to intrinsic variables such as $\hat{p}_0$ and $\hat{p}_i$ 
of the bicrossproduct basis. The new component of the momentum 
vector is referred to as simply $p_4$. When both constraints are 
satisfied, this component of the five-momentum is fixed at \beq 
p_4^2 = \ka^2 +\xi m^2;\eeq it's magnitude is quite large if 
$\ka$ is assumed to be near the Planck scale. The interpretation 
of $p_4$ is not addressed in this paper, although the lessons 
from standard SR suggest that it may have some physical 
significance. It's large magnitude, however, is expected to only 
have a very small effect on low-energy particle physics. 

Since the new five-dimensional space of momenta is flat, the 
addition (conservation) rules for these momenta should be linear 
just as in standard special relativity. Thus, all components of 
the five-momentum add linearly. It is important to note that 
momentum conservation in higher dimensions is a key feature of 
the proposed framework and sets it apart from some of the 
previous works; it is most similar to the proposal in 
\cite{GirelliLivine}. This feature also leads to some interesting 
consequences when scattering between many particles is considered 
in the next sections.

\section{Deformed Scalar Field Theory \label{s_scalar}}

This section reviews some concepts in standard quantum field 
theory of the scalar field and then applies these concepts to 
include the new momentum-space constraint. A prominent feature of 
this section and the next one is that the discussion does not 
begin with the postulation of an action or Lagrangian but rather 
with the definition of field expansions. 

\subsection*{Review} 
In standard scalar quantum field theory, a field $\phi$ is 
expressed as a superposition of modes in four-dimensional flat 
momentum space. Physical configurations are selected by imposing 
a mass-shell constraint. Thus a physical field configuration is 
\beq \label{phi1e} \begin{split} {\phi}(x) &= \int 
\frac{d^4p}{(2\pi)^4} \; {\phi}(p) e^{ip \cdot x} \; 
(2\pi)\delta^{(4)}(p^2-m^2)
\\  &= \int \frac{d^3p}{(2\pi)^3}\frac{1}{2p_0} \; 
{\phi}(p) e^{ip_0t} e^{i{\bf p}\cdot {\bf x}}. \end{split} \eeq 
To reach the second line, the integration $\int dp_0$ is carried 
out against the $\delta$-function and positive energy solutions 
are selected giving $p_0=+\sqrt{p_i^2+m^2}$. One sees that the 
invariant phase space integral for the field is \beq \int d\Pi = 
\int \frac{d^3p}{(2\pi)^3}\frac{1}{2p_0}. \eeq

The normalization of the functions ${\phi}(p)$ are fixed by an 
inner product defined by an integration in position space over a 
spatial hypersurface, \beq ({\phi}(x), {\phi}(x^\prime)) = i\int 
d^3x \; \left[ \phi^*(x)\pa_t \phi(x^\prime) - (\pa_t 
\phi^*(x))\phi(x^\prime)\right] = \delta^{(3)}(x-x^\prime). \eeq 
In momentum space, the integral is over spatial momenta, and the 
time derivative brings out a factor of energy on the right hand 
side. The resultant normalization is
 \beq ({\phi}(p), {\phi}(p^\prime)) = 
(2\pi)^3 \; p_0 \; \delta^{(3)}({\bf p}-{\bf p^\prime}), \eeq 
which is invariant under standard Lorentz transformations. The
inner product is positive-definite for solutions with $p_0>0$. 

The second-quantized version of the field is \beq {\phi}(x) = 
\int \frac{d^3p}{(2\pi)^3}\frac{1}{\sqrt{2p_0}}\; \left( a_{\bf 
p} e^{-ip_0t}e^{i{\bf p }\cdot{\bf x} } + a_{\bf p}^\da 
e^{ip_0t}e^{-i{\bf p }\cdot{\bf x} } \right), \eeq where 
$a^\da_{\bf p}$ and $a_{\bf p}$ are creation and annihilation 
operators obeying the usual algebra $[a_{\bf p},a^\da_{\bf 
p^\prime}] = (2\pi)^3\delta^{(3)}({\bf p }-{\bf p^\prime }).$ The 
propagator for the field is \beq \begin{split} {D}_F(x-y) = 
\langle 0| {\phi}(x){\phi}(y) |0\rangle &= \int 
\frac{d^3p}{(2\pi)^3} \frac{1}{2p_0} e^{-ip\cdot(x-y)} 
\\ &= \int \frac{d^4p}{(2\pi)^4} 
\frac{i}{p^2-m^2+i\epsilon} e^{-ip\cdot(x-y)}. \end{split}\eeq 
The calculation leading to the second line involves a particular 
but standard choice of integration contour along the $p_0$ 
direction. In momentum space, the propagator is written as \beq 
{D}_F(p) = \frac{i}{p^2-m^2+i\epsilon} \eeq and is very useful in 
computing matrix elements in perturbation theory.

\subsection*{Deformation}

In the deformed field theory, the fundamental momentum integral 
is five-dimensional and physical configurations of the scalar 
field are selected by imposing the two constraints 
(\ref{twoconstraints}). The field expansion is defined by \beq 
\phi(x,x_4) = \int \frac{d^5p}{(2\pi)^5} \; \phi(p,p_4) 
e^{ip\cdot x} e^{ip_4x_4}\; (2\pi)\delta^{(4)}(p^2-m^2)\; 
(2\pi)\delta^{(5)}(p^2-\xi p_4^2+\xi\ka^2). \eeq 

Integrating out first the $\ka$-shell $\delta$-function against 
$\int dp_4$ gives \beq \label{phi2e} \phi(x,x_4) = \int 
\frac{d^4p}{(2\pi)^4} \frac{1}{2p_4} \; \phi(p,p_4) e^{ip\cdot x} 
e^{ip_4x_4} \; (2\pi)\delta^{(4)}(p^2-m^2), \eeq with the $p_4$ 
momentum is now fixed at $p_4=\sqrt{\ka^2+\xi(p_0^2-p_i^2)}$. 
Note that resulting field expansion is similar to (\ref{phi1e}) 
with the four-dimensional measure replaced \beq 
\label{changemeasure} \int d^4p \ra \int \frac{d^4p}{p_4}.\eeq 
The latter is the integration measure for (anti)de-Sitter space 
of fixed radius $\ka$. Since a good integration measure should be 
positive and real, the factor $p_4$ should be positive and real 
as well. Consequently, the four dimensional integrals remaining 
in (\ref{phi2e}) appear to be restricted to the regime 
$\ka^2+\xi(p_0^2-p_i^2)>0. $ 

Integrating out the second delta-function and selecting the 
solutions with positive $p_0$, the field finally becomes \beq 
\phi(x,x_4) = \int \frac{d^3p}{(2\pi)^3} \frac{1}{2p_0} 
\frac{1}{2p_4} \; \phi(p) e^{ip_0t} e^{i{\bf p}\cdot {\bf 
x}}e^{ip_4 x_4}, \eeq where $p_0=\sqrt{p_i^2+m^2}$. With this 
substitution, the $p_4$ momentum becomes just a constant 
$p_4=\sqrt{\ka^2+\xi m^2}$ and since it was assumed that $m^2\ll 
\ka^2$, $p_4$ is real as required for the reality of the 
four-dimensional measure (\ref{changemeasure}). 

Very importantly, note that the four-dimensional integral does 
not appear to be restricted if the $\delta$-functions are removed 
in a different order. This suggests that the extra constraint 
does not really have a regulatory effect in the 4d sector as may 
at first appear from the discussion above. In any case, the 
result is that the natural phase space integral for the deformed 
field is \beq \label{dphase} \int d\Pi_\ka = \int 
\frac{d^3p}{(2\pi)^3}\frac{1}{2p_0}\frac{1}{2p_4}, \eeq which is, 
up to a constant, the same as in undeformed field theory. 
 

The normalization condition can be taken as \beq 
(\phi(p),\phi(p^\prime)) \propto (2\pi)^3 \; p_0 p_4 \; 
\delta^{(3)}({\bf p}-{\bf p^\prime}), \eeq which is positive 
definite in the selected sector where both $p_0$ and $p_4$ are 
positive. The quantization of the field can proceed as before by 
setting \beq \phi(x,x_4) = \int 
\frac{d^3p}{(2\pi)^3}\frac{1}{\sqrt{2p_0}\sqrt{2p_4}}\; \left( 
a_{\bf p} e^{-ip_0t}e^{i{\bf p }\cdot{\bf x} }e^{-ip_4x_4} + 
a_{\bf p}^\da e^{ip_0t}e^{-i{\bf p }\cdot{\bf x} }e^{ip_4x_4} 
\right) \eeq and imposing the same commutators $[a_{\bf 
p},a^\da_{\bf p^\prime}] = (2\pi)^3\delta^{(3)}({\bf p }-{\bf 
p^\prime })$ on the creation and annihilation operators. The 
propagator for the new field is \beq D_F(x_A-x_B) = \langle 0| 
\phi(x)\phi(y) |0\rangle = \int \frac{d^3p}{(2\pi)^3} 
\frac{1}{2p_0} \frac{1}{2p_4}\; e^{-ip\cdot(x-y)}e^{ip_4x_4}. 
\eeq This could also be written in terms of higher dimensional 
integrals as \beq \begin{split} D_F(x-y) &= \int 
\frac{d^4p}{(2\pi)^4} \left(\frac{i}{p^2-m^2+i\epsilon}\right) 
\frac{1}{2p_4} e^{-ip\cdot(x-y)}e^{ip_4x_4} 
\\ &= \int \frac{d^5p}{(2\pi)^5} 
\left(\frac{i}{p^2-m^2+i\epsilon}\right) 
\left(\frac{i}{p_4^2-\ka^2-\xi m^2+i\epsilon}\right) 
e^{-ip\cdot(x-y)}e^{-ip_4(x_4-y_4)}. \end{split} \eeq As before, 
these transformations involve particular choices of contour 
integrations, this time along the $p_0$ and $p_4$ directions. 

The propagator $D_F$ can also be written in momentum space using 
the last form above. For computations with Feynman rules, this 
propagator will be defined as \beq \label{DF1} D_F(p) = 
\frac{-1}{(p^2-m^2+i\epsilon)}\frac{1}{(p_4^2-\ka^2-\xi 
m^2+i\epsilon)}. \eeq The appearance of an additional factor in 
the denominator shows that the deformation introduces new poles 
into the propagator: In addition to the usual poles on the $p_0$ 
axis, the propagator also has new poles along the $p_4$ axis. The 
placement of the new poles on the axis of the extra dimension 
will become important in the discussion of the optical theorem 
and unitarity in interacting theories. 
 
Note that in defining this free theory, only those fields $\phi$ 
with positive $p_0$ and positive $p_4$ were selected. In 
principle, another set of scalar fields could be consistently be 
defined using the negative $p_4$ solutions to the constraint. In 
this paper, this sector of solutions is truncated, but it may 
still be interesting to explore it in more detail.

\subsection*{Interactions}

A deformed version of an interacting theory can be defined 
perturbatively through Feynman diagrams and scattering amplitudes 
$i\mc{M}_\ka$ can be computed by following a set of simple rules. 
Given a lack of an action from which to derive these rules, it 
seems reasonable to guess a set of rules and later check their 
consistency.

Every diagram has three main parts: external lines, internal 
lines, and vertices. The simplest rule dealing with the external 
lines is to do nothing, i.e. multiply the amplitude by $1$. As 
for the internal lines, the previous discussion suggests to write 
down the deformed propagator. The remaining parts of the Feynman 
rules deal with the vertices. To obtain an analog for $\phi^4$ 
theory, vertices in Feynman diagrams should be four-valent. The 
coupling constant can in principle be represented by any 
variable; it is chosen here as $-i\lambda$ in order to remain 
close to the notation of the undeformed theory. In accordance 
with the definition of the fields in five-dimensional flat 
momentum space, the momentum conservation $\delta$-function 
usually associated with vertices should also be five-dimensional. 
The argument of the $\delta$-function should be a sum of all the 
incoming fields' momenta in 5d. In case a Feynman diagram 
contrains a loop, the amplitude resulting from these rules should 
include an overall five-dimensional integral over this 
undetermined momentum. As usual, the amplitude may have to be 
divided by an appropriate symmetry factor. 

Using these Feynman rules, it is straight-forward to compute the 
amplitude for the first order scattering of two incoming scalar 
fields into two outgoing scalar fields. The amplitude is simply 
proportional to $\lambda$ so that $|\mc{M}_\ka|^2 = \lambda^2$ 
just like in standard $\phi^4$ theory. The amplitudes in the 
deformed and standard theories have the same form because there 
are no undetermined momenta, but the situation is not 
significantly more complicated for diagrams that contain loops, 
as is shown below.

\subsection*{Unitarity}

A prescription for writing transition amplitudes from Feynman 
diagrams as defined above can only be consistent if the resulting 
theory is unitary. This is because unitarity is the property that 
ensures time-reversal invariance and the conservation of 
probabilities. The following discussion of the optical theorem 
follows \cite{PeskinSchroeder} and checks that unitarity is 
preserved in $\phi^4$ theory to $\lambda^2$ order.

Consider a process whereby two particles with momenta $p,p_4$ and 
$p^\prime,p^\prime_4$ interact to produce another two particles 
with momenta $k,k_4$ and $k^\prime,k_4^\prime$. The $S$-matrix is 
defined as $S=1+iT$ where $T$ is related to the process amplitude 
$\mc{M}$ by \beq \langle k\, k_4,\, k^\prime\,k_4^\prime|T|p\, 
p_4,\, p^\prime\,p_4^\prime \rangle =  i\mc{M}( p\, p_4,\, 
p^\prime\, p_4^\prime \ra k\, k_4,\, k^\prime\, k_4^\prime)\; 
(2\pi)^5 \, \delta^{(5)}(p+p^\prime-k-k^\prime). \eeq Note that 
conservation condition of the fifth component of momentum, 
$p_4+p_4^\prime-k_4-k_4^\prime=0$, is automatically satisfied 
because all these components are fixed to $\sqrt{\ka^2+\xi m^2}$ 
by the on-shell conditions. A theory is unitary if the $S$-matrix 
satisfies $SS^\da= 1$; in terms of the $T$-matrix, this 
translates into \beq \label{opt1} -i(T-T^\da) = T^\da T. \eeq 

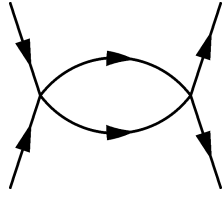
\begin{figure}[!t] 
\begin{center}
\begin{fmffile}{sclopp}     
  \fmfframe(1,7)(1,7){       
   \begin{fmfgraph*}(100,70) 
    \fmfleft{i1,i2}          
    \fmfright{o1,o2}
    \fmflabel{$p$}{i1}
    \fmflabel{$p^\prime$}{i2}
    \fmflabel{$k$}{o1}
    \fmflabel{$k^\prime$}{o2}
    \fmf{fermion}{i1,v1}
    \fmf{fermion}{i2,v1}
    \fmf{fermion,left=0.5,tension=0.2,label=$q$}{v1,v2}
    \fmf{fermion,right=0.5,tension=0.2,label=$q^\prime$}{v1,v2}
    \fmf{fermion}{v2,o1}
    \fmf{fermion}{v2,o2} 
    \fmffreeze
   \end{fmfgraph*}
  }
\end{fmffile}
\end{center} \caption{\label{f_scloop} Loop diagram in $\phi^4$ theory.}
\end{figure}

The unitarity condition can be checked to first non-trivial order 
in perturbation theory by considering the diagram in Figure 
\ref{f_scloop}. In the `center of mass' frame, the incoming 
particles have momenta $(p_0,p_i,p_4)$ and $(p_0,-p_i,p_4)$ so 
that the total momentum is $(2p_0,0,2p_4)$. The energy $p_0$ can 
be as large as desired, but $p_4$ is fixed by the on-shell 
conditions. Therefore, the matrix element $\mc{M}$ for this 
process can be taken to be a function of the energy $p_0$ alone, 
$\mc{M}=\mc{M}(p_0).$ The requirement that $\mc{M}$ be real for 
low energies leads by standard arguments to the conclusion that 
\beq \mathrm{Disc} \,\mc{M}(p_0) = 2i\, \mathrm{Im}\, 
\mc{M}(p_0+i\epsilon), \eeq relating the discontinuity in the 
amplitude due to the $+i\epsilon$ prescription, to the imaginary 
part of the amplitude. This is useful for checking unitarity 
because $\mathrm{Im}\, \mc{M}(p_0+i\epsilon)$ also appears on the 
left hand side of (\ref{opt1}).

Checking the optical theorem for the loop diagram in Figure 
\ref{f_scloop} involves computing the discontinuity of the 
amplitude and then relating that result to the right hand side of 
(\ref{opt1}). The amplitude is \beq \begin{split} & 
i\mc{M}\;\delta^{(5)}(p+p^\prime-k-k^\prime)  \\ & \qquad = 
\frac{\lambda^2}{2} \int \frac{d^5q}{(2\pi)^5} \int 
\frac{d^5q^\prime}{(2\pi)^5} \; \left( f_0 f_4 \right) \; 
(2\pi)^5 
\delta^{(5)}(p+p^\prime-q-q^\prime)\delta^{(5)}(q+q^\prime-k-k^\prime), 
\end{split}\eeq where \beq \begin{split} f_0 &= 
\left(\frac{1}{q^2-m^2+i\epsilon}\right) \left(\frac{1}{q^{\prime 
2}-m^2+i\epsilon}\right), 
\\ f_4 &= \left(\frac{1}{q_4^2 -\ka^2-\xi m^2 + i\epsilon}\right) 
\left( \frac{1}{q_4^{\prime 2} -\ka^2-\xi m^2+i\epsilon}\right). 
\end{split}\eeq
The integration is actually over one undetermined five-momentum 
propagating around the loop, but it is written in terms of 
integrals $d^5q$ and $d^5q^\prime$ and some $\delta$-function for 
later convenience. Functions $f_0$ and $f_4$ contain the 
propagators of the internal particles. A useful feature of the 
model is that all the dependence (except for the overall 
constants) on the deformation parameter $\ka$ and the extra 
dimensions $q_4$ and $q_4^\prime$ can be neatly put in the 
function $f_4$.

The discontinuity in the resulting expression can be computed 
using the usual cutting rules algorithm 
\cite{PeskinSchroeder,Cutkosky}. In the case of the diagram in 
Figure \ref{f_scloop}, the cuts should be made  along the two 
internal lines. The propagators for each of the lines consist of 
the factors in the function $f_0$ and $f_4$. Thus the algorithm 
replaces all the factors in the denominator by $\delta$-functions 
as follows \beq \begin{split} \frac{1}{(q^2-m^2+i\epsilon)} &\ra 
-2\pi i \delta(q^2-m^2) \\ \frac{1}{(q_4^2-\ka^2-\xi m^2 
+i\epsilon)} &\ra -2\pi i\delta(q_4^2-\ka^2-\xi m^2)\end{split} 
\eeq and similarly for the primed variables. The 
$\delta$-functions allow to evaluate the integrals along the 
$q_0, q_0^\prime, q_4, q_4^\prime $ directions, giving the 
discontinuity as \beq \label{optres1} \mathrm{Disc} \, \mc{M} = 
\frac{\lambda^2 }{2} \int \frac{d^3q}{(2\pi)^3} \frac{1}{2q_0} 
\frac{1}{2q_4} \; \int 
\frac{d^3q^\prime}{(2\pi)^3}\frac{1}{2q_0^\prime}\frac{1}{2q_4^\prime} 
\; (2\pi)^5 \delta^{(5)}(p+p^\prime-q-q^\prime). \eeq

Returning to the right hand side of (\ref{opt1}), the product 
$TT^\da$ can be split by inserting an identity operator. This 
gives \beq \begin{split} & \langle k\, k_4,\, 
k^\prime\,k_4^\prime|T^\da T|p\, p_4,\, p^\prime\,p_4^\prime 
\rangle \\ & \qquad\;\; = \sum_n \left(\prod^n \int 
\frac{d^3q}{(2\pi)^3}\frac{1}{2q_0}\frac{1}{2q_4} \right) \langle 
k\, k_4,\, k^\prime\,k_4^\prime|T^\da|\{q\, q_4\}\rangle \langle 
\{q\, q_4\}|T|p\, p_4,\, p^\prime\,p_4^\prime \rangle, 
\end{split}\label{TT2} \eeq where the sum is over all possible 
intermediate configurations and the factors in parenthesis are 
the one-particle phase space integrals from (\ref{dphase}) (the 
index characterizing the different particles is omitted in order 
not to confuse it with the components of the momenta, see 
\cite{PeskinSchroeder} for more details on notation). For the 
whole expression to be of order $\lambda^2$, each of the matrix 
elements $\langle \cdot|T|\cdot\rangle$ on the right hand side 
should correspond to just single-vertex diagrams valued 
$\lambda$. There is only one possible such configuration and it 
contains two intermediate particles. Thus (\ref{TT2}) reduces to 
\beq \label{optres2} \begin{split} & \langle k\, k_4,\, 
k^\prime\,k_4^\prime|T^\da T|p\, p_4,\, p^\prime\,p_4^\prime 
\rangle \\ & \qquad\;\; = \lambda^2 \; \int 
\frac{d^3q}{(2\pi)^3}\frac{1}{2q_0}\frac{1}{2q_4} \; \int 
\frac{d^3q^\prime}{(2\pi)^3}\frac{1}{2q_0^\prime}\frac{1}{2q_4^\prime} 
\; (2\pi)^5 \delta^{(5)}(p+p^\prime-q-q^\prime), \end{split} \eeq 
which is very close to the result (\ref{optres1}). The usual
discrepancy of $1/2$ is due to the fact (\ref{optres2}) should be 
symmetrized with respect to the intermediate particles labelled 
by $q,q_4$ and $q^\prime, q^\prime_4$. 

The result shows that the proposed deformed Feynman rules 
generate amplitudes that satisfy the optical theorem, at least to 
second order in the coupling constant $\lambda$. The deformed 
theory is therefore unitary and consistent.

\subsection*{Comments}

This section ends with some comments regarding the role of the 
fifth component of momentum in the deformed $\phi^4$ theory. 

Since every external particle must have its extra momentum 
component set to $p_4=\sqrt{\ka^2+\xi m^2}$, the overall momentum 
conservation $\delta$-function for any two-particle to 
two-particle process is trivially satisfied for this component. 
Moreover, this rigidity in the $p_4$ component implies that any 
process that takes two incoming particles to four or more 
outgoing particles is kinematically forbidden. In standard 
$\phi^4$ theory, such processes are allowed, so the deformed 
$\phi^4$ theory presented in this section makes some very 
different predictions from the standard theory at higher orders 
in the perturbation expansion. 

An interesting question to ask at this point is whether Feynman 
diagram calculations at higher orders make sense in the deformed 
theory. In other words, is the deformed theory renormalizable? To 
answer this question properly, one should first understand the 
physical meaning of the fifth dimension in momentum space. Some 
aspects of this question are raised again in the next section, 
after having introduced a deformed version of quantum 
electrodynamics.

\section{Deformed Electrodynamics \label{s_QED}}

In this section, the deformation prescription is applied to 
quantum electrodynamics. A set of deformed Feynman rules for 
interacting fermions and photons are given that preserve 
diagrammatic Ward identities even in the presence of the new 
scale. A matrix element for a phenomenologically interesting 
scattering process is calculated.

\subsection*{Properties of the Fifth Momentum Component}

Conservation of momentum in the fifth, extra, dimension can 
potentially have a great effect on kinematics. The primary goal 
of a deformed QED theory is to reproduce all the amplitudes of 
standard quantum electrodynamics at low energies (and, hopefully, 
to generate some new effects). Therefore, conservation of the 
momentum in the fifth-component should be compatible with all the 
standard QED processes - this condition can provide some 
interesting information about the properties of the new component 
of momentum, despite it not having a satisfying physical 
interpretation at the moment.

In the section on scalar fields, $p_4$ was always positive and 
was set to a constant $\sqrt{\ka^2+\xi m^2}$ for on-shell 
particles. When deforming quantum electrodynamics, a theory with 
many types of particles, several questions arise about the 
properties of the $p_4$ momentum. Should all particle types have 
the same magnitude of the deformation constant $\ka$, or should 
the deformation scale $\ka$ depend on the particle type? Should 
physical particles all have their $p_4$ momenta carry the same 
sign, or should some have $p_4>0$ while others have $p_4<0$? To 
answer such questions, it is useful to consider some simple 
examples of test scattering experiments in order to extract the 
properties the fifth component of the momentum that are 
consistent with observations.

As a first test experiment, consider the reaction $e^+e^-\ra 
\tau^+\tau^-$. In the center of mass frame, the incoming 
particles both have $p_4$ of a magnitude $\sqrt{\ka_e^2+\xi 
m_e^2}$; the constant $\ka_e$ has an $e$ subscript to emphasize 
that it is labelling the $\ka$-shell of an electron. Suppose 
further that the outgoing particles both have their fifth 
components of momentum of magnitude $\sqrt{\ka_\tau^2+\xi 
m_\tau^2}.$ If momentum is conserved, $p_4^{e^+}+p_4^{e^-} = 
p_4^{\tau^+}+p_4^{\tau^-}$, then one can infer that if all the 
momenta are of the same sign, the values of $\ka_e$ and 
$\ka_\tau$ must be different. If, on the other hand, the momenta 
for the electron/positron and taon/anti-taon pairs have opposite 
signs, then the conservation equation is satisfied trivially. 

As another test example, consider a process $e^+e^- \ra 
\tau^+\tau^-\tau^+\tau^-$ or any similar one in which there are 
more than two outgoing particles. Such processes cannot occur if 
all the $p_4$ have the same sign. However, momentum conservation 
in the fifth component is again trivially satisfied if particles 
and anti-particles carry $p_4$ of opposite signs. It should be 
concluded, therefore, that $p_4$ must be different for particles 
and anti-particles. No conclusion can be reached on the values of 
$\ka$ for the different particle species. In the work that 
follows, however, the values for $\ka$ for all the species are 
set equal for simplicity.

As a final example, consider the process $e^+ \ra e^+ 
\gamma\cdots\gamma$ where an electron radiates one or several 
photons. It should be noted that this kind of process is 
forbidden in standard QED in vacuo; it can occur, however, in the 
presence of an external electric field, for example when the 
electron travels in a medium such as a liquid. Assuming that the 
deformation scale $\ka_\gamma$ for the photon is non-zero, one 
can also deduce that the radiation process is forbidden due to 
momentum conservation in the fifth dimension. It would be 
interesting to consider what happens to this process in the 
presence of an external electric field, but the following 
discussion, however, is concentrated only on in-vacuo processes 
for simplicity. Nonetheless, in order to distinguish the photon 
from the fermions, its deformation scale is non-zero and is 
labelled by $\ka_\gamma =\mu$. The possibility that $\mu=0$ is 
not considered here.

To summarize, the test scattering experiments considered suggest 
that particles and anti-particles should have the same 
deformation scale $\ka$ but opposite signs of $p_4$; the photon 
can be said to carry a positive $p_4$ momentum, and its 
deformation scale is called $\mu$. There is no indication as to 
whether the values of $\ka$ for different fermion species should 
be the same or different, but for simplicity, the following 
discussion assumes that all fermions have the same $\ka$. Given 
these insights, the rest of this section deals with a deformed 
model of QED which implements these features.

\subsection*{Feynman Rules}

The definition of QED can be split into three parts: the free 
fermion field, the free electromagnetic field, and the 
interactions between them. To start with, the deformation of the 
free fields can be carried out in a similar fashion as done to 
the scalar field in the previous section. The field expansions 
for the fermion fields are taken to be \beq 
\begin{split} \psi(x,x_4) &= \int 
\frac{d^3p}{(2\pi)^3}\frac{1}{\sqrt{2p_0}}\frac{1}{\sqrt{2p_4}} 
\sum_s \left( a_\pv^s u^s(p) e^{-ip\cdot x}+ b^{s\da}_\pv v^s(p) 
e^{ip\cdot x }\right)\, e^{-ip_4 x_4}, \\ \overline{\psi}(x,x_4) 
&= \int 
\frac{d^3p}{(2\pi)^3}\frac{1}{\sqrt{2p_0}}\frac{1}{\sqrt{2p_4}} 
\sum_s \left( b_\pv^s \overline{v}^s(p) e^{-ip\cdot x}+ 
a^{s\da}_\pv \overline{u}^s(p) e^{ip\cdot x }\right)\, e^{ip_4 
x_4}.\end{split} \eeq The operators $a,b,$ their conjugates, and 
the four-dimensional spinors $u^s,v^s$ and their conjugates are 
the usual ones. The deformation is seen in the new oscillating 
functions $e^{\pm ip_4 x_4}$ and in new factors of $p_4^{-1/2}$. 
Note that the operators $b^{s\da}$ and $a^{s\da}$ create 
particles with the same four-momentum $p$ but with opposite 
values of the fifth momentum component $p_4$, in accordance with 
the insights obtained earlier. The fermion propagator can be 
computed as in the section on the scalar field; it should be \beq 
\frac{\fs{p}+m}{p^2-m^2+i\epsilon} \ra 
\left(\frac{\fs{p}+m}{p^2-m^2+i\epsilon}\right)\left(\frac{1}{ 
p_4^2-\ka^2-\xi m^2 + i\epsilon}\right). \eeq 

The deformation of the electromagnetic field $A(x,x_4)$ can also 
be written in analogy to the above result for fermions. The 
photon propagator should thus be \beq 
\frac{\eta_{\mu\nu}}{q^2+i\epsilon} \ra 
\left(\frac{\eta_{\mu\nu}}{q^2+i\epsilon} \right) \left( 
\frac{1}{q_4^2-\mu^2+i\epsilon}\right). \eeq The main difference 
between this deformation and the one used for the fermion 
propagator is that here the deformation parameter is $\mu$ 
instead of $\ka$.

For the purposes of introducing an interaction between fermions 
and the photons, it is useful to have a Lagrangian or action 
formulation of the theory. From the structure of the fermion 
propagator, it is possible to guess the action \beq S_\psi 
\propto \int \frac{d^4p}{(2\pi)^4} \int \frac{dp_4}{(2\pi)} \; 
\big[ \overline{\psi}(-p,-p_4)\left(p_4^2-\ka^2 
\right)\left(\fs{p}-m\right)\psi(p,o_4)\big].\eeq The dimension 
of this action can be made to match the standard one by 
introducing a new dimensionless variable $\ti{p}_4 = p_4/\ka$ and 
rewriting the action as \beq \label{Spsidef} S_\psi = \int 
\frac{d^4p}{(2\pi)^4} \int \frac{d\ti{p}_4}{(2\pi)} \; \big[ 
\overline{\psi}(-p,-\ti{p}_4)(\ti{p}_4^2-1)(\fs{p}-m)\psi(p,\ti{p}_4) 
\big]. \eeq Some factors of $\ka$ are absorbed into the fields 
$\psi$ and $\overline{\psi}$ in this step. The possible criticism 
that this action leads to a higher-derivative theory when 
expressed in a five dimensional position-space representation 
together with the usual problems associated with such theories, 
is postponed to the comments at the end of this section. 

Similarly, one can guess an action for the photon field that 
could give rise to the desired propagator. In terms of a 
dimensionless momentum $ \ti{q}_4  = q_4/\mu$, such an action 
could be, in Lorentz gauge, \beq \label{SAdef} S_A = 
\frac{1}{2}\int\frac{d^4q}{(2\pi)^4} \int 
\frac{d\ti{q}_4}{(2\pi)}\; \big[ 
A_\mu(-q,-\ti{q}_4)(\ti{q}_4^2-1)(q^2)A^\mu(q,\ti{q}_4) \big]. 
\eeq 

To complete the definition of deformed QED, one should define the 
interaction vertices and the coupling constant. The interaction 
term is usually obtained from the action by minimally coupling 
the fermion field to the photon field via a replacement of 
partial derivatives by covariant derivatives, $\pa_\mu \ra 
\pa_\mu + e A_\mu$. In momentum space, this replacement roughly 
translates to $p_\mu \ra p_\mu + eA_\mu(q)$ and a condition 
imposing momentum conservation at the vertex. In the deformed 
scenario, the interaction term is taken to be \beq \label{Sint}
\begin{split} S_{int} &= e \left(\int \frac{d^4p\, d\ti{p}_4}{(2\pi)^5} \right) 
\left(\int \frac{d^4p^\prime\, d\ti{p}_4^\prime}{(2\pi)^5} 
\right) \left(\int \frac{d^4q\, d\ti{q}_4}{(2\pi)^5} \right) \; 
\\ 
 &\qquad\qquad \times \Big[(\ti{p}_4^2-1)^{1/2} \overline{\psi}(p,p_4)\Big]\,\Big[ 
(\ti{q}_4^2-1)^{1/2} \gamma^\mu A_\mu(q,\ti{q}_4)\Big]\,\Big[ 
(\ti{p}_4^\prime-1)^{1/2}\psi(p^\prime,\ti{p}_4^\prime)\Big] \\ & 
\qquad\qquad \times (2\pi)^5 \; 
\delta^{(4)}(p+p^\prime+q)\;\delta(\ka \ti{p}_4 + 
\ka\ti{p}_4^\prime +\mu \ti{q}_4).
\end{split} \eeq
Despite the long form, this interaction term is not difficult to 
understand. The first line expresses that three particles are 
interacting in five dimensions. The factor $e$ in front of the 
integrals is the usual coupling constant used in perturbation 
series. The second line identifies these particles as two 
fermions and a photon. There are also some factors depending on 
the dimensionless fifth component of momentum. These factors can 
be understood as arising from (\ref{Spsidef}) by the minimal 
coupling prescription. The factor associated with the $A$ field 
is introduced rather ad-hoc here. It can be thought of as a 
modification to the minimal coupling definition, as a factor that 
makes all fields enter symmetrically, or it can be justified 
a-posteriori since it simplifies some expressions in computations 
of transition amplitudes. The third line of (\ref{Sint}) imposes 
momentum conservation in five-dimensions. The second 
$\delta$-function contains $\ka$ and $\mu$ factors because 
momentum conservation is imposed on the original dimensionful 
variables. 

The remaining part of defining Feynman rules is related to 
labelling of external lines in a diagram. Again, the guiding 
factor for guessing the Feynman rules is for them to reproduce 
the usual amplitudes as closely as possible. For this purpose, a 
suggested set of rules is to give each external line the usual 
QED factor and further divide by a factor $(\ti{p}_4-1)^{1/2}$. 
This factor serves to cancel a similar factor that is introduced 
via the vertex coupling. 

Because of the feature that the dependence on momenta in the 
extra dimension can always be factored away from the dependence 
on the other momenta, all amplitudes can be expressed as products 
of two pieces, \beq \mc{M}_\ka = \mc{M}_s\,\mc{M}_d. \eeq The 
first factor $\mc{M}_s$ is the `standard' piece containing the 
usual propagators, coupling constants, and integrals over 
four-momenta. The second factor $\mc{M}_d$ is the `deformation' 
piece where factors related to the new component of momentum can 
be found.

\subsection*{Examples}

\begin{figure}[!t] 
\begin{center}
\begin{fmffile}{epem}    
  \fmfframe(1,7)(1,7){          
   \begin{fmfgraph*}(100,70)    
    \fmfleft{i1,i2}    
    \fmfright{o1,o2}
    \fmflabel{$p_A$}{i1}
    \fmflabel{$p_A^\prime$}{i2}
    \fmflabel{$p_B$}{o1}
    \fmflabel{$p_B^\prime$}{o2}    
    \fmf{fermion}{i1,v1}
    \fmf{fermion}{v1,i2}
    \fmf{photon,label=$q$}{v1,v2}
    \fmf{fermion}{o1,v2}
    \fmf{fermion}{v2,o2} 
    \fmffreeze
   \end{fmfgraph*}
  }
\end{fmffile}
\end{center} \caption{\label{f_epem} Electron-position scattering.}
\end{figure}
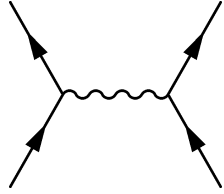

As a simple example of the application of the deformed Feynman 
rules, consider the scattering process in Figure \ref{f_epem}. 
Suppose that the incoming particles are an electron/positron 
pair, and that the outgoing particles are a muon/anti-muon pair. 

The five momenta of the electron and positron, in the center of 
mass, are $p_A=(p_0,p_i,p_4)$ and $p_A^\prime=(p_0,-p_i,-p_4)$. 
In this frame, the five-momentum of the photon is completely 
determined at $(2p_0,0,0)$. The resultant amplitude-squared for 
this process, $|\mc{M}_{\ka}|^2_{\,avg}$, can be obtained using 
the usual spin-averaging relations. It can be written separating 
the usual amplitude $\mc{M}$ from the corrections as follows \beq 
\label{Mdefee} |\mc{M}_{\ka}|^2_{\,avg} = 
\frac{1}{4}\sum_{spins}|\mc{M}_\ka|^2 = 
\frac{1}{4}\sum_{spins}|\mc{M}_s|^2 |\mc{M}_d|^2. \eeq Here 
$\mc{M}_s$ is the usual amplitude of QED, while $\mc{M}_d$ is a 
deformation factor that depends on the extra dimension and knows 
about the scale $\ka$. In $\mc{M}_d$, the deformation factor in 
the photon propagator is cancelled by two $(\ti{q}_4^2-1)^{1/2}$ 
factors coming from the interaction vertices. Also, the 
contributions of such vertex factors to the fermions are 
cancelled by the external line rule. So the part of the amplitude 
that could depend on the scale $\ka$ is actually unity, 
$\mc{M}_d=1.$ Overall, then, the deformed amplitude-squared 
$|\mc{M}_{\ka}|^2_{\,avg}$ agrees exactly with the amplitude of 
standard QED. Nonetheless, the observed cross-sections in the two 
theories do not necessarily have to be the same. This issue is 
explained in more detail in the comments at the end of this 
section.

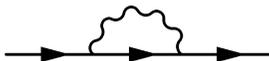
\begin{figure}[!th] 
\begin{center}
\begin{fmffile}{eself}    
  \fmfframe(1,7)(1,7){          
   \begin{fmfgraph*}(100,70)    
    \fmfleft{i1}    
    \fmfright{o1}
    \fmflabel{$p_A$}{i1}
    \fmf{fermion}{i1,v1,v2,o1}
    \fmffreeze
    \fmf{photon,left,tension=0.1,label=$q$}{v1,v2}
    \fmffreeze
   \end{fmfgraph*}
  }
\end{fmffile}
\end{center} \caption{\label{f_eself} Electron self-energy.}
\end{figure}

More significant deviations from standard QED start to appear in 
diagrams that contain loops, such as the electron self-energy 
diagram shown in Figure \ref{f_eself}. The amplitude for this 
diagram is \beq \label{Mdefself} \mc{M}_\ka = \mc{M}_s \left( 
\int \frac{d\ti{q}_4}{(2\pi)}\; 1 \right). \eeq The standard 
piece $\mc{M}_s$ is divergent as usual. The deformation piece, 
written out explicitly in terms of an integral over the fifth 
momentum dimension, is clearly divergent as well. Therefore, it 
turns out that the deformed amplitude for the diagram is more 
problematic than in the standard theory. 

It is tempting to hope that the simple form of the new infinite 
factor could be renormalized away in some way. In fact, one can 
see that divergent factors such as in (\ref{Mdefself}) appear for 
every loop in a diagram; thus, the divergences could be removed 
by introducing an additional Feynman rule with the effect of 
either multiplying every loop in a diagram by a normalized 
function of the $p_4$ component of the loop momentum, or putting 
the virtual particles in the loop on the $\ka$-shell. These seem 
ad-hoc at the moment but are perhaps an interesting avenue for 
further investigations.

\subsection*{Comments}

The deformation factors that appear in the amplitudes are all 
very simple: they are either unity or integrals of unity over an 
undetermined component of momentum in the extra dimension. One 
can understand the source of this feature by defining new derived 
fields $\Psi(p,p_4) = (\ti{p}_4^2-1)^{1/2} \psi(p,p_4)$ for the 
fermions and similarly for the gauge field. In terms of these 
variables, the action of the deformed electrodynamics looks 
exactly the same as that of standard QED with the only exception 
that all integrals are five dimensional instead of four; these 
extra integrals are the sources of the new divergences. Writing 
the action in terms of the derived fields is also helpful in 
seeing why the higher-order actions (\ref{Spsidef}) and 
(\ref{SAdef}) do not cause problems for the interacting theory: 
interactions are defined only in the specific sector of the 
theory corresponding to the well-behaved derived fields. (Indeed, 
these properties follow almost `by construction' from the minimal 
coupling postulate.)

It should be stressed that despite the close resemblance of the 
deformed theory, expressed in terms of the new derived fields, to 
standard QED, these two theories are not equivalent. The main 
difference between them is the postulate of momentum conservation 
in five dimensions in the deformed version, which, since $p_4$ is 
constant for on-shell particles, effectively introduces a new 
`charge' or quantity that has to be conserved. Recall that in the 
scalar field theory of the previous section, conservation of 
momentum in the extra component had the effect of setting to zero 
all amplitudes of processes taking two incoming particles to four 
or more outgoing particles. Similar consequences could also arise 
in QED. A possible place where these effects could arise is in 
the radiation processes $e^-\ra e^- \gamma \cdots \gamma$ 
mentioned at the beginning of this section. Deviations from the 
standard results would appear in diagrams with many vertices but 
would not depend on the energy of the incoming particle. Such 
behavior provides the possibility that the deformation could be 
tested, measured, and even disproved in high precision 
experiments.

Another reason why this deformed model is non-trivial goes back 
to the original motivation for introducing the extra dimension, 
which is to implement curvature in momentum space. It may be that 
the physically relevant momenta are ones that, like the 
bicrossproduct coordinates on de Sitter space, satisfy modified 
dispersion relations 
\cite{Liberati,GirelliLivine,correctionsloops}. In this case, 
then to convert the amplitudes calculated in the deformed theory 
to observed cross-sections, one should replace all the 
five-dimensional momentum variables used in the calculations by 
four-dimensional coordinates on the physical space. The simple 
amplitude (\ref{Mdefee}) would give a deformed cross section 
$\sigma_\ka$ that deviates slightly from the usual $\sigma_s$, 
\beq \sigma_\ka = \sigma_s \left(1+ c_1 \frac{\ti{p_0}}{\ka} + 
\ldots \right), \eeq where $c_1$ is some order-unity constant and 
the other terms are suppressed by higher powers of $\ka$. 
Computations of scattering processes in this model would then 
resemble the methods used in previous work on process thresholds 
in DSR \cite{JudesVisser,Heyman}. At the moment, however, a 
complete and satisfying explanation for the physical meaning of 
the various momentum variables is missing, so it would be very 
useful to obtain experimental feedback on this matter.

\section{Discussion \label{s_discussion}}

In this work, a simple prescription for deforming field theories 
is introduced to incorporate a new energy scale $\ka$ while 
explicitly preserving Lorentz invariance. The motivation for 
writing such deformed theories is the need to study quantum 
gravity phenomenology. The proposed deformed models have some 
features that are desirable for such studies, but they also 
suffer from some weaknesses and ambiguities.

One of the main positive features of the proposed deformed model 
is that a new scale $\ka$ is introduced into the usual field 
theory framework while keeping the usual ability to 
perturbatively compute amplitudes of scattering processes. This 
is achieved by introducing an auxiliary dimension together with a 
new constraint to momentum space. (The essential features of the 
deformation are almost the same whether the resulting curvature 
in momentum space is positive or negative.) The move to a higher 
dimension is argued to be analogous to the shift from space to 
space-time that is a result of adopting special relativity. Two 
important consequences of this formalism are that Lorentz 
symmetry is explicitly preserved and that momentum conservation 
(in the higher-dimensional space) is linear as usual. 

Field theories are constructed as deformations of standard 
$\phi^4$ theory and standard quantum electrodynamics. The scalar 
field theory is explicitly shown to be unitary to second order in 
the coupling parameter. The deformed version of QED is argued to 
preserve the usual diagrammatic Ward identities (which are 
usually a direct consequence of gauge-invariance) and reproduce 
the usual scattering amplitudes to leading order. Some 
possibilities for seeing discrepancies between the deformed 
models and the standard theories are also discussed. The most 
promising effect stems out of the new momentum conservation rules 
that can suppress some higher-than-second-order interactions. 
Another effect, which is however dependent on the definition of 
the cross section and the choice of observable momentum 
variables, scales as an energy over the constant $\ka$. The 
possibility of such a modification has also been found in other 
works on scattering in the presence of a new length scale 
\cite{Hossenfelder:2003jz}. 

With regards to deformed quantum electrodynamics, the requirement 
that process amplitudes are consistent at leading order with the 
usually calculated quantities points to the interesting 
conclusion that particles and antiparticles carry momentum of 
opposite signs in the extra dimension. This observation is both 
surprising and encouraging since it implies that some properties 
of the new extra dimension, which in this work is treated as a 
byproduct of modelling quantum gravity, can be deduced from low 
energy experiments.

The proposed deformation model admittedly also has some 
weaknesses. Conceptually, one would like to have a physical 
interpretation for the fifth dimension, both in momentum space 
and the dual position space. Such an interpretation is missing 
and is not attempted in this paper, albeit it does seem to be 
related to mass. Other weaknesses of the proposed model are 
related to the specific deformed field theories discussed. Both 
of the deformed interacting field theories are defined 
perturbatively through a set of Feynman rules - it would be 
satisfying to have a formulation of the theories in terms of an 
action principle. Despite some comments along these lines in the 
section on quantum electrodynamics, one would like to have more 
control over this matter, particularly in the domain of 
quantization of the gauge field. With regards to the scattering 
amplitudes, the new divergences arising in loop diagrams should 
be understood in more detail and the question of 
renormalizability should be addressed systematically. As 
mentioned earlier, this issue may be intimately connected with 
the interpretation of the new extra dimension. It should be 
checked that the potential effects are not in contradiction with 
the results of existing high-precision experiments.

A lot of work remains to be done on Planck-scale phenomenology. 
In the context of the deformation proposal advocated in this 
work, future work should address and answer the weaknesses 
outlined above, as well as make concrete predictions for 
observations in real experiments. It would also be interesting to 
understand how the proposed approach is related, if at all, to 
other works on field theory in the context of Deformed Special 
Relativity \cite{Daszkiewicz,Arzano,Dimitrijevic}, other field 
theories in curved momentum space \cite{Kadyshevsky}, other 
models with invariance scales \cite{Stability}, or  
space-time-matter theory which also uses an extra dimension 
\cite{STmatter}.

\vspace{0.3cm} {\bf Acknowledgments.$\;\;\;$} I would like to 
thank F. Girelli, S. Hossenfelder, J. Kowalski-Glikman, S. Majid, 
R. Mann, F. Markopoulou, R. Myers, M. Pospelov, and L. Smolin for 
many useful discussions. I would also like thank members of the 
Department of Applied Mathematics and Theoretical Physics, 
University of Cambridge, for their comments on a talk based on 
this work.

\end{document}